# Role of Ni, Si and P on the formation of solute-rich clusters under irradiation in Fe-Cr alloys

P-M. Gueye[a], B. Gómez-Ferrer[a], C. Kaden[b,] C. Pareige[a,*]

[a] *UNIROUEN, INSA Rouen, CNRS, Groupe de Physique des Matériaux, 76000 Rouen, France*

[b] *Institute of Resource Ecology, Helmholtz-Zentrum Dresden - Rossendorf, Dresden, 01328, Germany*

*Corresponding Author:    Cristelle Pareige

Email:              cristelle.pareige@univ-rouen.fr

Phone Number:       +33 (0)2-32-95-51-31

Postal Address:     GPM - UMR 6634 CNRS
                    Rouen Normandy University
                    Avenue de l'Université - BP12
                    76801 St. Étienne du Rouvray
                    FRANCE

# Abstract

After irradiation of Fe-Cr alloys of low purity (model alloys of F-M steels), minor solute elements as P, Ni and Si have been shown to create solute clusters which significantly contribute to hardening and might be associated with small dislocation loops. In order to understand the role of each impurity on the formation of the nano-features formed under irradiation and the eventual synergies between the different species, Fe-15at.%Cr-X (X=Si, Ni, P, NiSiP) alloys of different composition have been ion irradiated and characterized using atom probe tomography. Irradiation were performed at 300 °C up to 2.5 dpa in four alloys: Fe15CrNi, Fe15CrSi, Fe15CrP and Fe15CrNiSiP. Influence of C atoms implanted during irradiation on the nanostructure evolution is also discussed. The study of the evolution of the nanofeatures formed under irradiation with the dose as a function of the composition highlights the role of P and C on the formation of the nano-clusters and confirm the radiation-induced nature of solute-rich clusters.

# 1  Introduction

The operation temperature window of ferritic/martensitic steels under irradiation is limited at high temperature by irradiation creep and at low temperature by radiation embrittlement. Part of the radiation embrittlement is due to radiation hardening caused by the formation of intragranular nanometric features. Two families of nanofeatures contribute to the hardening: (i) dislocation loops [1–3] and (ii) clusters and/or precipitates formed either by radiation-induced or by radiation-enhanced mechanisms [4–7]. These clusters and/or precipitates are often constituted by a variety of nano-objects of different composition and structure, each of these objects having its own contribution to the hardening [8–11]. In low-Cr Fe-Cr model alloys of ferritic/martensitic steels, minor elements as P, Ni and Si have been shown to impact the dislocation loop distribution (decrease in size and increase of number density) [3] and create solute-rich clusters (SRC) which significantly contribute to the hardening [9,11]. The fact that the SRCs are observed in alloys with Ni, Si and P at impurity levels, the bell-shape[1] dependence of their solute content with temperature [4,12,13] and the systematic positive RIS (radiation induced segregation) of Ni, Si and P at point defect sinks [14] are strong indications that their formation is induced by the irradiation and not thermodynamically enhanced. Comparison of the number density of TEM invisible dislocation loops obtained by Object Kinetic Monte Carlo (OKMC) simulations by Chiapetto et al. [15] with the number density of the SRCs measured by atom probe tomography (APT) [4] strongly suggested that these objects are the same features and that the SRCs are associated to small point defect clusters. Recent OKMC simulations by Castin et al. [16] confirmed this hypothesis and highlighted the dominant role of the immobilized self-interstitial atoms clusters in the nucleation of the solute-rich clusters (SRCs). Since migrating P can be trapped by substitutional P atoms creating very stable complexes [17,18], it could be possible that these stable complexes act as a nuclei for the formation of the SRCs.

In order to understand the role of each minor element (namely Ni, Si and P) on the formation of the SRCs and the possible synergies between these species, but also to confirm the nature of the SRCs, four Fe-15at.%Cr-X (X=Si, Ni, P, NiSiP) alloys of different composition have been ion irradiated at 300

---

[1] At very low temperature point defects having very low mobility, no dragging of the solute towards the point defect sinks occurs. At high temperature back-diffusion is important removing concentration gradients. As a consequence, segregation is stronger at intermediate temperatures.

°C up to 2.5 dpa and characterized using atom probe tomography. The 3D distribution maps of the chemical species have been analyzed to characterize the evolution of the cluster´s characteristics, i.e. size, number density, composition and size distribution, with the dose and the minor element composition.  Cr also forms clusters at the higher dose but treatment of these clusters will not be addressed in this paper. The Cr concentration being the same in all the alloys, differences in behaviour cannot be attributed to the formation of Cr-rich clusters.

## 2 Materials and experiments

### 2.1 Materials

Four different Fe-15at.%Cr-X model alloys (X = Ni, Si, P, NiSiP) (Table 1) were cast by OCAS in an induction vacuum furnace. From each lab cast, the pieces were heat treated in a pre-heated furnace at 1200°C for 1 h 30 min, subsequently hot rolled and finally air-cooled down to room temperature obtaining a fully ferritic microstructure (for further details see [19]). Composition provided in Table 1 was measured by APT and isotopic overlaps were systematically corrected.  The grain size of the alloys is very similar: 239 µm for Fe15CrNiSiP, 253 µm for Fe15CrP, 217 µm for Fe15CrSi and 240 µm for Fe15CrNi [19].

Table 1– Intragranular compositions of the Fe-15at.%Cr-X (X = NiSiP, Ni, Si, P) model alloys measured by APT in at%. The uncertainties corresponds to the maximum of the difference between average concentration of X and the concentration of X measured in the different APT volumes.

| Short Name | Cr (at%) | Ni (at%) | Si (at%) | P (at%) | Al (at%) | C (at%) | N (at%) |
|---|---|---|---|---|---|---|---|
| Fe15CrNiSiP (G394) | 14.8 ± 0.4 | 0.098 ± 0.012 | 0.42 ± 0.05 | 0.065 ± 0.024 | 0.056 ± 0.004 | 0.011 ± 0.014 | 0.03 ± 0.02 |
| Fe15CrNi (G391) | 15.55 ± 0.11 | 0.099 ± 0.001 | 0.013 ± 0.013 | 0.003 ± 0.001 | 0.070 ± 0.001 | 0.010 ± 0.004 | 0.08 ± 0.02 |
| Fe15CrSi (G392) | 15.06 ± 0.18 | 0.006 ± 0.001 | 0.422 ± 0.019 | 0.002 ± 0.001 | 0.068 ± 0.001 | 0.006 ± 0.02 | 0.02 ± 0.03 |
| Fe15CrP (G393) | 15.0 ± 0.7 | 0.006 ± 0.01 | 0.016 ± 0.012 | 0.042 ± 0.002 | 0.064 ± 0.001 | 0.002 ± 0.001 | 0.01 ± 0.02 |

### 2.2 Irradiation conditions

The samples (8-10 x 10 x 1 mm$^3$) were mirror polished before irradiation using mechanical polishing in several steps down to diamond suspension of 1 µm. Subsequently, the deformed surface layer was removed by low-temperature (~0°C) electropolishing using a solution of 98% of ethylen glycol monobutyl ether and 2% of perchloric acid.

The model alloys were irradiated with 5 MeV Fe$^{2+}$ ions at 300°C using a 3MV-Tandetron accelerator at the Ion Beam Center of HZDR, Dresden, Germany. The irradiation was designed in such a way that the nominal displacement damage was reached at a depth of 500 nm in order to avoid surface effect and the influence of the implantation peak [20]. The displacement damage and implantation profiles given in Figure 1 were calculated using SRIM (Stopping and Range of Ions in Matter) [21,22]  with the ''Quick Kinchin-Pease calculation" mode following the recommendation of Stoller et al. [23] and using a displacement threshold of 40 eV [24]. In order to guaranty a homogeneous lateral exposure over the whole set of samples the focussed ion beam was scanned over the area of the samples during the irradiation. The scanning frequencies were 1041 Hz and 1015 Hz in horizontal and vertical direction, respectively. The ion flux was measured continuously by means of Faraday cups (current measurement) and integrated to obtain the ion fluence. The ion flux was $1.3 \cdot 10^{11}$ $cm^{-2} \cdot s^{-1}$ corresponding to a damage rate of $5 \cdot 10^{-5}$ dpa/s at 500 nm depth. The irradiation temperature of 300°C was maintained by fixing the samples on a heating target. The temperature control was based on a thermocouple placed on the backside of one sample. The irradiation doses reached at 500 nm are 0.1, 0.5 and 2.5 dpa. Results obtained at 0.1 dpa for all alloys and at 0.5 dpa for Fe15CrNiSiP are already published in [25].

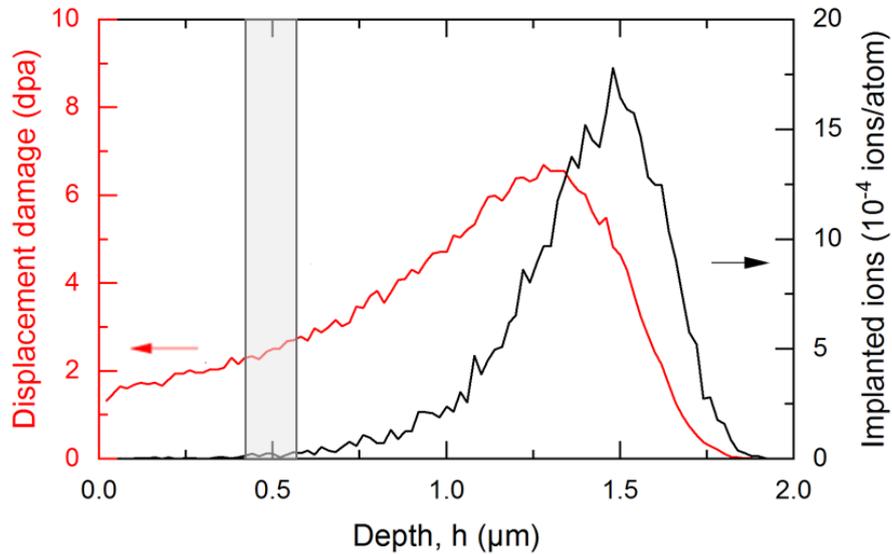

Figure 1: Displacement damage and injected Fe concentration profiles for the target dose 2.5 dpa at 500 nm with 5 MeV Fe-ions. In grey, the region investigated by APT.

## 2.3 Atom probe technique and data analysis methods

The APT samples of irradiated materials were prepared by lift-out of a chunk and subsequent annular milling using a Scanning Electron Microscope – Focus Ion Beam (SEM-FIB ZEISS Crossbeam X540). At least three APT tips from at least two different grains have been analysed for each condition. All the tips were prepared at a depth of (500 ± 100) nm from the surface in order to investigate the regions were the target doses are reached i.e. 0.1 dpa, 0.5 dpa or 2.5 dpa. The final milling was performed with a Ga beam energy of 2 kV in order to reduce implantation of Ga ions in the material. The maximum concentration of Ga found in the samples did not exceed 0.05 at%.

APT analyses of the alloys were conducted using a high-resolution local electrode atom probe, LEAP 4000 XHR (CAMECA), with a detector efficiency of 36%. The samples were set at a temperature of 55 K in order to avoid preferential evaporation of Cr atoms. The specimens were electrically pulsed with a pulse fraction of 20% of the DC voltage with a pulse repetition rate of 200 kHz. The detection rate was set up between 0.1% and 0.35%. 3D reconstructions were performed using IVAS software (CAMECA). The reconstruction factors have been adjusted for every analysed sample (i.e. every APT volume). The compression factor $\xi$ has been derived from the crystallographic angles between poles observed on the desorption maps and the field factor ($k$) has been derived from the expected interplanar distance at the poles. The values used for $\xi$ and $k$ have ranged between 1.5-1.65 and 3.6 – 5 (for tips on micro-coupons), respectively. APT samples of non-irradiated materials were electropolished. In that case, the field factor was between 2.1 and 3.3.

The data treatment was performed using the GPM APT data software for atom probe users developed by GPM, Rouen, France. The distribution of the minor solute elements (Ni, Si and P) in the matrix before and after irradiation has been quantified using a statistical tool based on the first nearest-neighbour distance distributions (1NN method) [26]. An iso-concentration method (ICM) [12,27] for cluster identification has also been applied when clusters are clearly visible and can be identified by the tool. In this paper, only clusters of minor solute elements are characterized.

### 2.3.1 1NN method

The frequency distributions of the 1NN distances of homo-pairs (X-X) being either Ni, Si or P—have been measured experimentally for every APT volume in regions far away from field desorption lines and poles. To build the histograms the sampling size used is 0.05 nm. In order to increase statistics, averaging has been done over all the analysed volumes for a given irradiation condition for each alloy. The experimental frequency distributions have been smoothed by averaging over 3 successive points and normalized to 100 (frequencies are thus in percentage). In case of random distribution of atoms, the probability density $P(r)$ to find two 1NN X-X atoms at a distance $r$ within $dr$ for diluted elements is given by [26]:

$$P^{1NN}(r) = 4\pi Q C_0\, r^2 \exp\left(-\frac{4}{3}\pi Q C_0 r^3\right) \qquad (1)$$

Where $Q$ is the detection efficiency of APT and $C_0$ is the nominal concentration per unit of volume in the analysed volume.

Figure 2 presents an example of experimental 1NN P-P frequency distribution (green curve) in an irradiated Fe15CrNiSiP alloy compared with the 1NN random distribution of the same concentration of P (black curve). The subtraction of the random frequency distribution from the experimental one reveals two peaks (red curve): one associated with the clusters and one associated with the matrix. The position of these two peaks depends on the concentration of the phases. The higher the concentration, the shorter the P-P distances. In the following, this curve will be named $\Delta_{1NN}$.

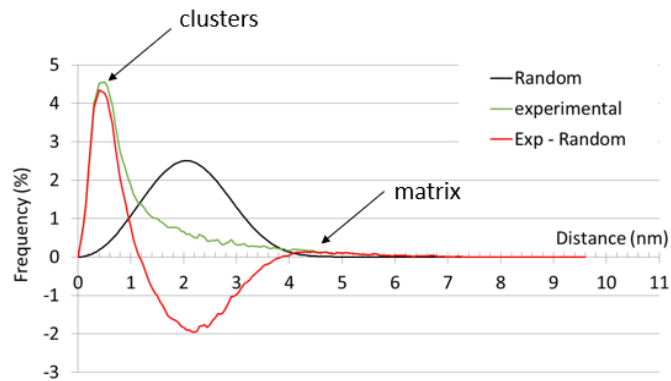

Figure 2: Experimental 1NN P-P frequency distribution (green curve) in an irradiated Fe15CrNiSiP alloy compared with the 1NN random distribution for the same P concentration (black curve). The red curve corresponds to the difference between the experimental P-P curve and the random 1NN curve ($\Delta_{1NN}$). The peak at short P-P distance is associated to P clusters. The peak at large distance corresponds to P-P pairs in the matrix.

### 2.3.2 Cluster characterization

The iso-concentration method (ICM) is a cluster identification procedure based on three parameters [27–29]: a concentration threshold ($C_{th}$), a minimum number of atoms ($N_b$) detected in the clusters and a distance *d* corresponding to the distance below which two atoms belong to the same cluster (d = 0.5 nm). The selection of the parameters $C_{th}$ and $N_b$ is made in such a way that ghost clusters (i.e. selection of stochastic fluctuations in the matrix) are avoided in a random solid solution of the same composition. $C_{th}$ ensures that less than 0.01% of the filtered atoms in the random solution would have this concentration [27]. Calculations are performed on 3D concentration maps built from sampling of APT volumes in small boxes of 1 nm in size [28]. $N_b$ is chosen in order to identify zero clusters in the random solution for the selected $C_{th}$ value. These parameters were estimated for each APT volume. The important point is to ensure that the choice made allows the correct identification (in size and shape) of all the clusters visible on the 3D images. $C_{th}$ depends on the concentration in solutes. It varies from 1.2 to 3.2 at.% and $N_{min}$ varies from 6 to 10 [25].

The number density ($N_V$) of the identified clusters was determined by a simple ratio of the number of the observed clusters to the overall analysed volume. After application of cluster identification methods, clusters are often found to be surrounded by a matrix shell rich in Fe and Cr which are erroneously associated to the clusters. These shells are systematically removed using an erosion method before measuring the size. The radius of each cluster was calculated as indicated in equation (2) considering the clusters as spherical:

$$R_{clus} = \sqrt[3]{\frac{3nV_{at}}{4\pi Q}} \qquad (2)$$

with *n* the number of detected atoms in each cluster, $V_{at}$ the Fe atomic volume ($a_0^3/2$ with $a_0$ the lattice parameter) and *Q* the detector efficiency. This equivalent radius is calculated for every cluster and the average radius, R, is finally calculated. To measure the composition of the clusters, the following procedure was applied: erosion profiles were drawn by class size and atoms belonging to the core of the clusters were selected (i.e. atoms from the plateau so that the interface is not considered). This guarantees a correct measurement of the composition of the clusters [28,30]. In order for the plateau to be visible, it is important to adapt the step size used to draw the profiles. A step size of 0.03 nm has been used. The mass spectrum is calculated with the selected atoms in order to calculate the composition of the clusters. Isotopic overlaps are corrected and background noise removed. The composition values were averaged over all the APT analysed volumes for a given condition.

## 3 Results

The 3D distributions of P, Si and Ni in Fe15CrP, Fe15CrSi, Fe15CrNi and Fe15CrNiSiP irradiated up to 2.5 dpa are presented Figure 3. At 0.1 dpa, only P maps reveal clustering. At 2.5 dpa, clusters are visible on P, Ni and Si maps. At this dose, comparison of Ni maps in Fe15CrNi and Fe15CrNiSiP shows a stronger clustering when Ni is associated to P and Si. This is revealed by the fact that more clusters are visible on the elemental map when the three species are altogether (Figure 3c). 1NN and $\Delta_{1NN}$ distributions at 0.1 dpa are reported in Figure 4 for homo-pairs and in Figure 5 for hetero-pairs. When segregated dislocation lines are observed in the APT volumes, they are removed in order to only probe 1NN distributions associated to segregation at point defect clusters or dislocation loops. Comparison of $\Delta_{1NN}$ curves at 0.1 dpa and before irradiation is provided in Figure 6. Before irradiation, slight deviations from the random distribution are observed. After irradiation, these deviations increased. Two clear maxima appear on the $\Delta_{1NN}$ curves revealing the presence of clusters embedded in the diluted matrix. As already mentioned in [25], whereas no clustering is visible on the spatial distribution maps of Ni and Si at 0.1 dpa, the 1NN and $\Delta_{1NN}$ curves show the presence of P, Ni and Si clustering. Figure 5 shows that Ni, Si and P form clusters altogether.

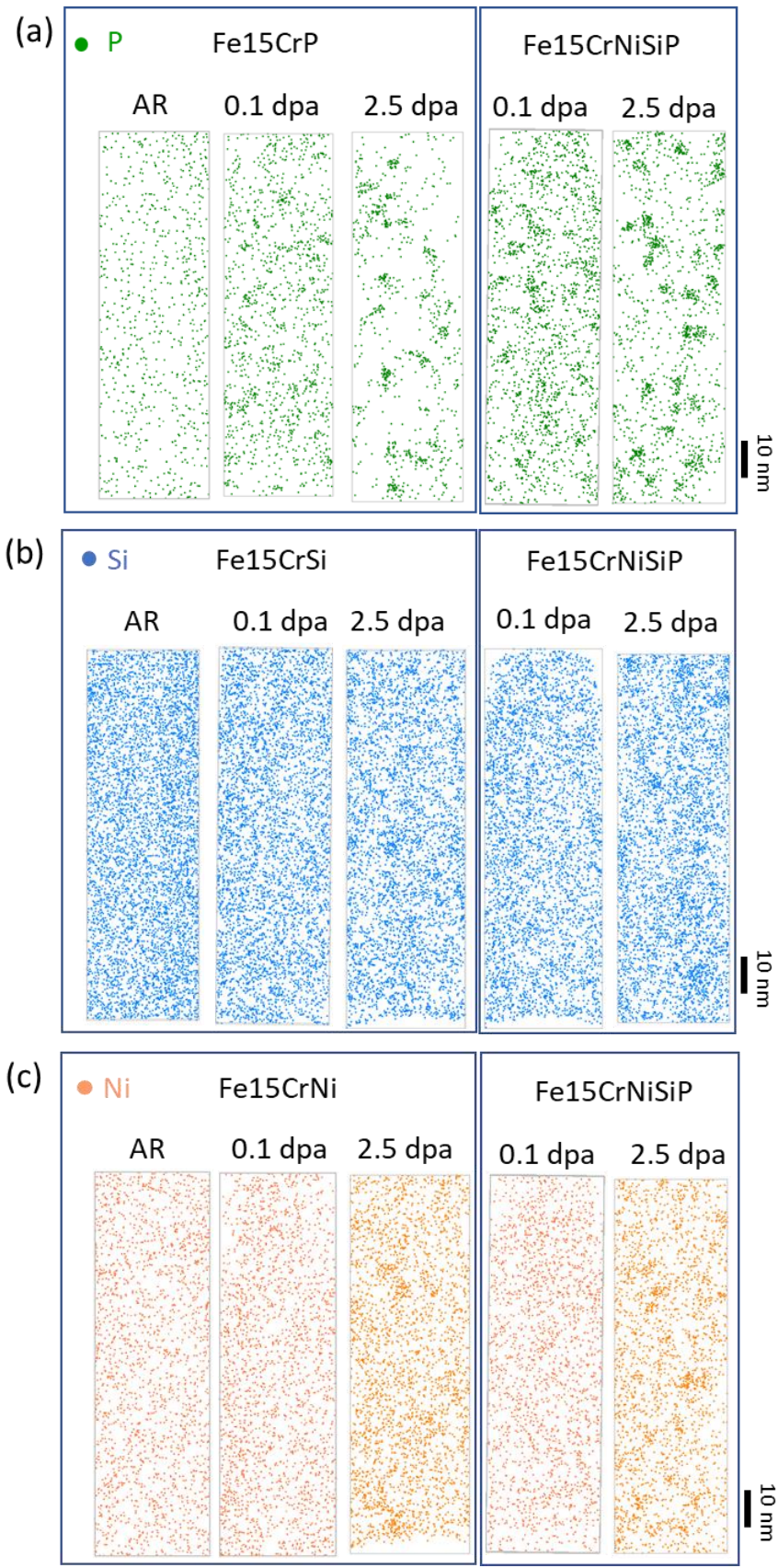

Figure 3: 3D-distribution maps of P (green), Ni (orange) and Si (blue) in ion irradiated Fe15CrP, Fe15CrSi, Fe15CrNi and Fe15CrNiSiP before irradiation (AR), at 0.1 dpa, 0.5 dpa and 2.5 dpa at 300 °C.

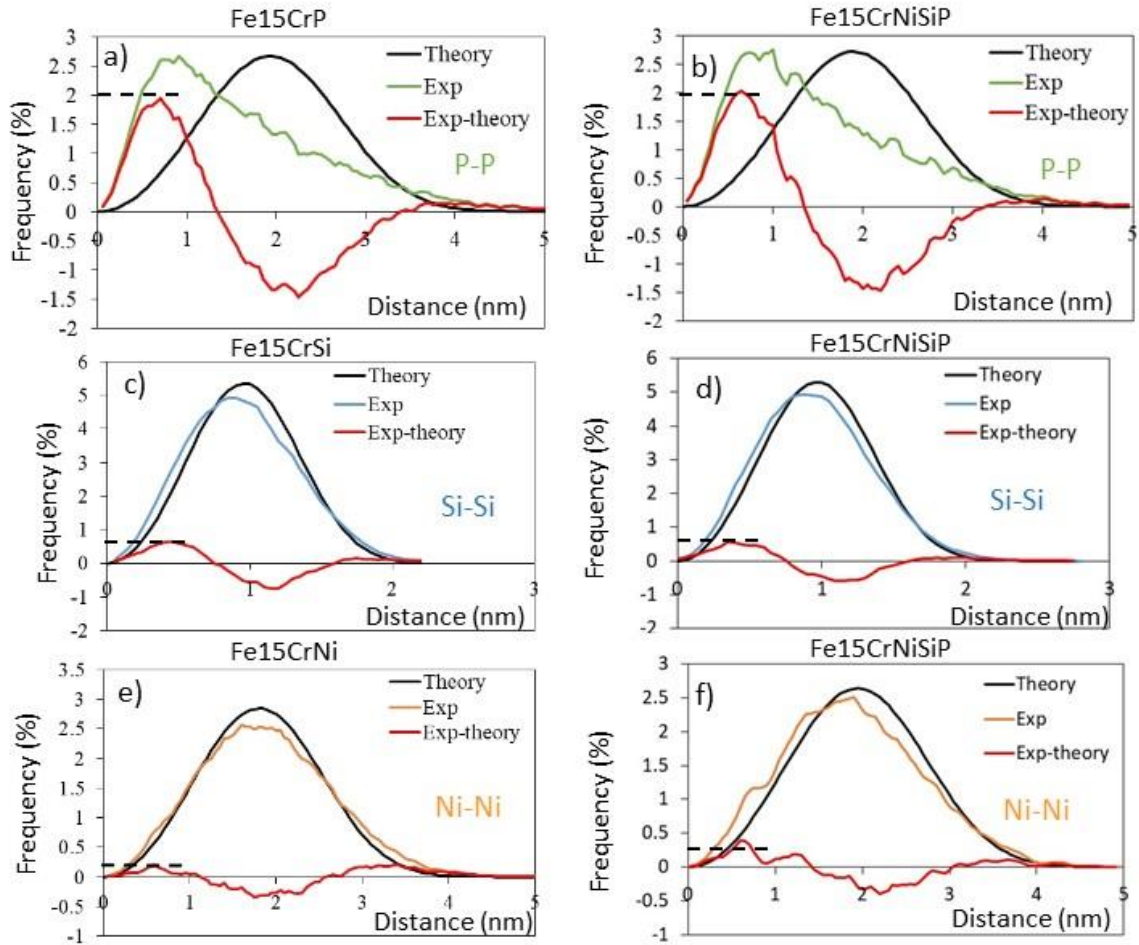

Figure 4: Experimental and theoretical 1NN distributions and $\Delta_{1NN}$ curve at 0.1 dpa at 300 °C for P-P pairs in (a) Fe15CrP and (b) Fe15CrNiSiP, Si-Si pairs in (c) Fe15CrSi and (d) Fe15CrNiSiP, Ni-Ni pairs in (e) Fe15CrNi and (f) Fe15CrNiSiP. The black distribution corresponds to the random distribution of atoms (binomial distribution). The green, blue and orange distribution are the experimental 1NN distributions of P-P, Si-Si and Ni-Ni, respectively. The red curve is the $\Delta_{1NN}$ curve. The dotted lines correspond to the maximum level observed in alloys with only one specie and are reported on the graphs obtained for the alloy with the three species for comparison.

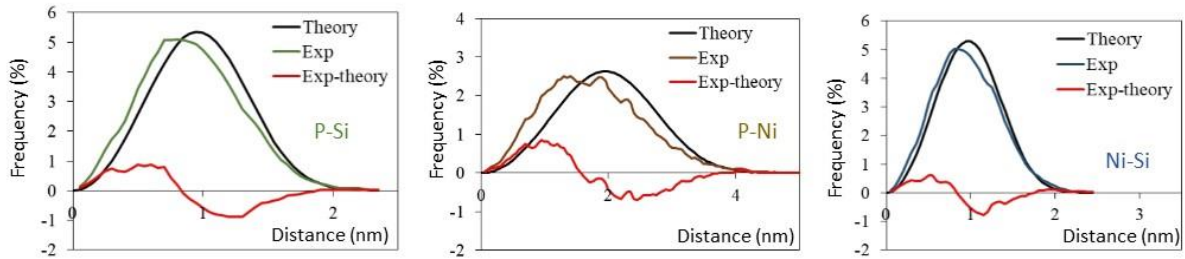

Figure 5: Experimental and theoretical 1NN distributions and $\Delta_{1NN}$ curve at 0.1 dpa for (a) P-Si pairs, (b) P-Ni pairs and (c) Ni-Si pairs in Fe15CrNiSiP

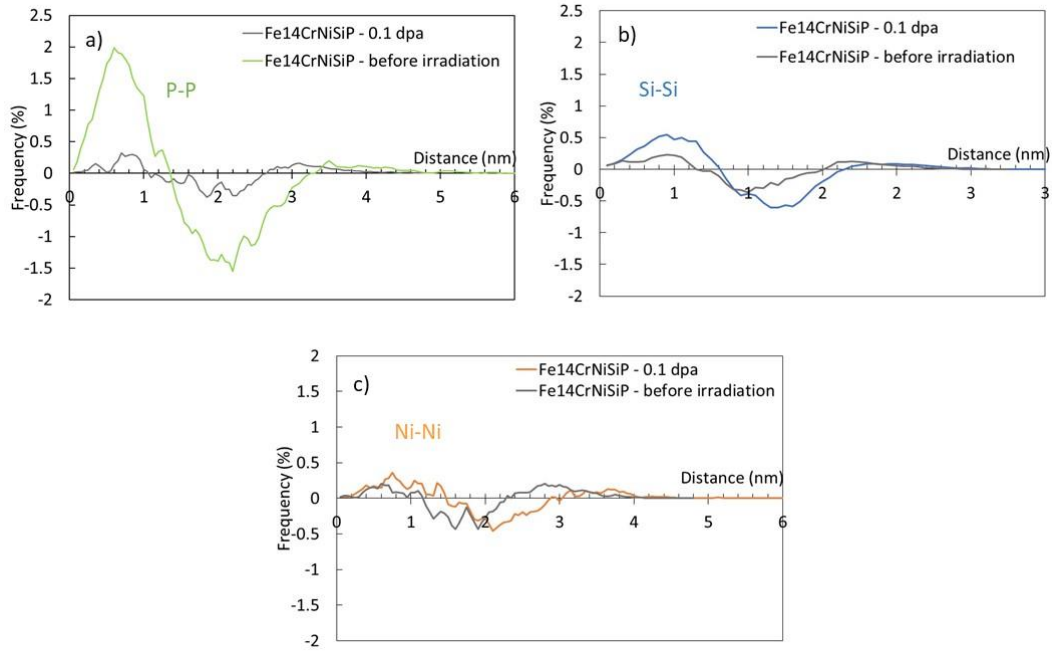

Figure 6: $\Delta_{1NN}$ curves for (a) P-P pairs, (b) Si-Si pairs and (c) Ni-Ni pairs in Fe15CrNiSiP before irradiation and at 0.1 dpa at 300 °C.

The 1NN distributions of P-P, Ni-Ni and Si-Si pairs at 2.5 dpa in all alloys are reported in Figure 7. Comparison of $\Delta_{1NN}$ curves in alloys with only one minor element to the alloys with the three of them shows that clustering of Ni, Si and P is stronger in alloys containing Ni+Si+P altogether: the amplitude of the peaks corresponding to the clusters is greater in the alloy containing the three minor solutes. Table 2 presents the evolution of the SRCs composition in Fe15CrNiSiP with the dose. Whereas P concentration is very high at 0.1 dpa, one notes a decrease at 0.5 and 2.5 dpa, whereas Ni and Si concentrations increase. At 2.5 dpa, dislocation loops are also visible on the 3D images (Figure 8). They are detected with the same filtering criteria as those used to detect SRCs ($C_{th}$, Nmin and d). Dislocation loops were identified by visual inspection of the individual objects i.e. when a hole is visible in the feature detected by ICM, it is considered as a dislocation loop. The size of each loop has been measured considering the outer limit of the segregated region, not the dislocation loop by itself as the dislocation loop is not visible with APT. Consequently, the size is probably overestimated. The number density, radius and composition are provided in Table 2. Only the largest loops are expected to be visible with TEM.

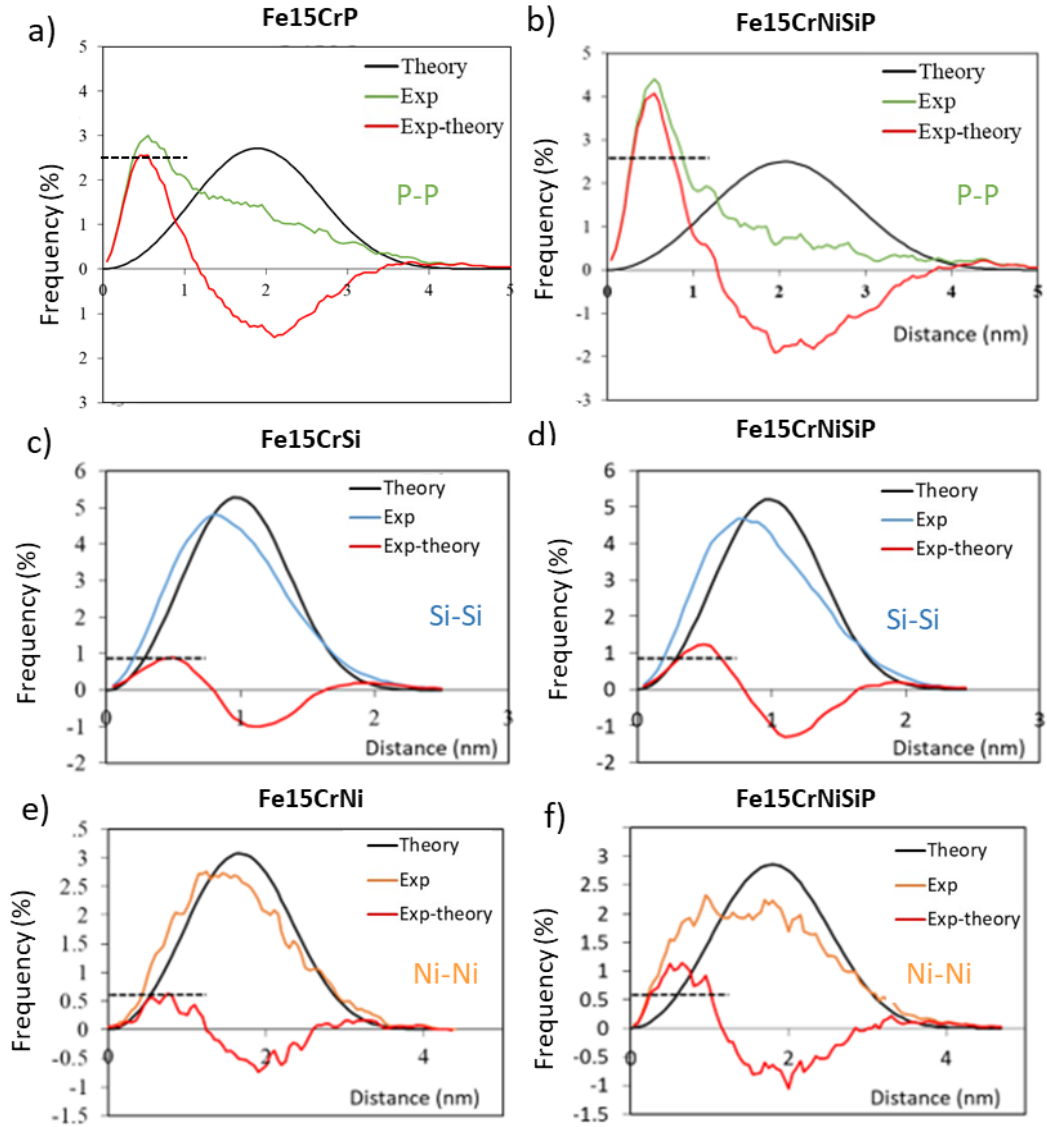

Figure 7 : 1NN and $\Delta_{1NN}$ distributions at 2.5 dpa for P-P pairs in (a) Fe15CrP and (b) Fe15CrNiSiP, Si-Si pairs in (c) Fe15CrSi and (d) Fe15CrNiSiP, Ni-Ni pairs in (e) Fe15CrNi and (f) Fe15CrNiSiP. The black curves correspond to the 1NN distribution of X-X pairs in the corresponding random alloys. Dashed-lines mark the maximum $\Delta_{1NN}$ values measured in alloys with only one element. The dotted lines correspond to the maximum level observed in alloys with only one specie and are reported on the graphs obtained for the alloy with the three species for comparison.

Table 2 : Composition (in at.%), radius and number density of SRCs and dislocation loops in Fe-14Cr-NiSiP alloys irradiated up to 2.5 dpa.

| Short Name | | Cr | Ni | Si | P | C | Number density ($10^{23}$ m$^{-3}$) | Radius (nm) |
|---|---|---|---|---|---|---|---|---|
| 0.1 dpa [25] | SRCs | 17 ± 3 | 1.1 ± 0.8 | 0.8 ± 0.8 | 7.0 ± 2.0 | - | 3.0* ± 0.9 | 0.8 ± 0.2 |
| 0.5 dpa [25] | SRCs | 17 ± 3 | 2.7 ± 0.6 | 4.0 ± 1.1 | 2.8 ± 1.3 | - | 6.4 ± 1.9 | 1.1 ± 0.3 |
| 2.5 dpa | SRCs | 18.4 ± 0.4 | 3.0 ± 0.2 | 2.7 ± 0.2 | 2.7 ± 0.2 | 0.54 ± 0.07 | 3.5 ± 0.3 | 1.6 ± 0.6 |
| | Dislocation loops | 17.8 ± 0.7 | 2.5 ± 0.2 | 2.5 ± 0.2 | 2.2 ± 0.2 | 0.7 ± 0.2 | 1.4 ± 0.2 | 2.8 ± 0.7 |

\* only clusters detected by the iso-concentration method i.e. whose radius is larger than 0.6 nm.

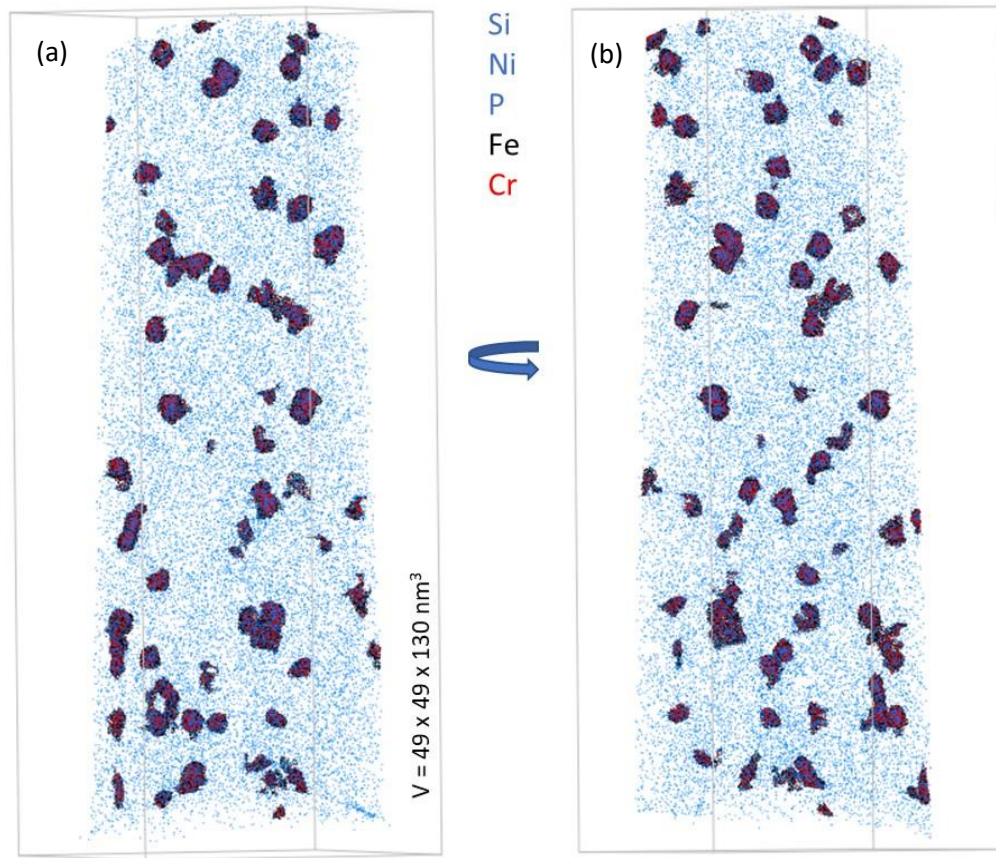

Figure 8: Fe15CrNiSiP irradiated at 2.5 dpa. Both SRCs and dislocation loops are visible. All Si, Ni and P atoms are presented. Fe and Cr atoms in SRCs are visible (concentration threshold used: Ni+Si+P > 3 at.%). Difference between (a) and (b) is only the orientation of the volume. Depending on the orientation of the loops in the volume, loops can appear as SRC in the image. The hole in the middle of the segregated zone is not always visible. A video is available in supplementary materials.

Table 3 gives the C concentration before and after irradiation at 2.5 dpa. One can note a significant increase in C content after irradiation. Even if a share of the C concentration can be assigned to the FIB preparation (used to prepare irradiated sample, but not AR samples which were electropolished), the measured C concentration is mainly due to C contamination during irradiation. It appears that some carbon contamination is unavoidable in spite of the efforts to limit it [31,32]. This is confirmed by Figure 9 which presents the 3D distribution maps of Si, Ni and C in Fe15CrSi and Fe15CrNi alloys irradiated at 2.5 dpa. These images together with 1NN and $\Delta_{1NN}$ curves for C-Si and C-Ni pairs (Figure 9) reveal the co-segregation of Ni and Si with C, but do not exclude that Ni or Si or C form some clusters alone. The clusters of C atoms cannot come from FIB preparation which implies a homogeneous distribution of C. C clusters were not observed in Fe15CrNiSiP and Fe15CrP. We do not have any explanation for this observation at this stage.

Table 3: C concentration before irradiation (AR) and at 2.5 dpa.

|  | Fe15CrP | Fe15CrSi | Fe15CrNi | Fe15CrNiSiP |
|---|---|---|---|---|
| AR | 0.0020 ± 0.0003 | 0.006 ± 0.02 | 0.010 ± 0.009 | 0.01 ± 0.01 |
| 2.5 dpa | 0.07 ± 0.07 | 0.15 ± 0.04 | 0.11 ± 0.04 | 0.26 ± 0.04 |

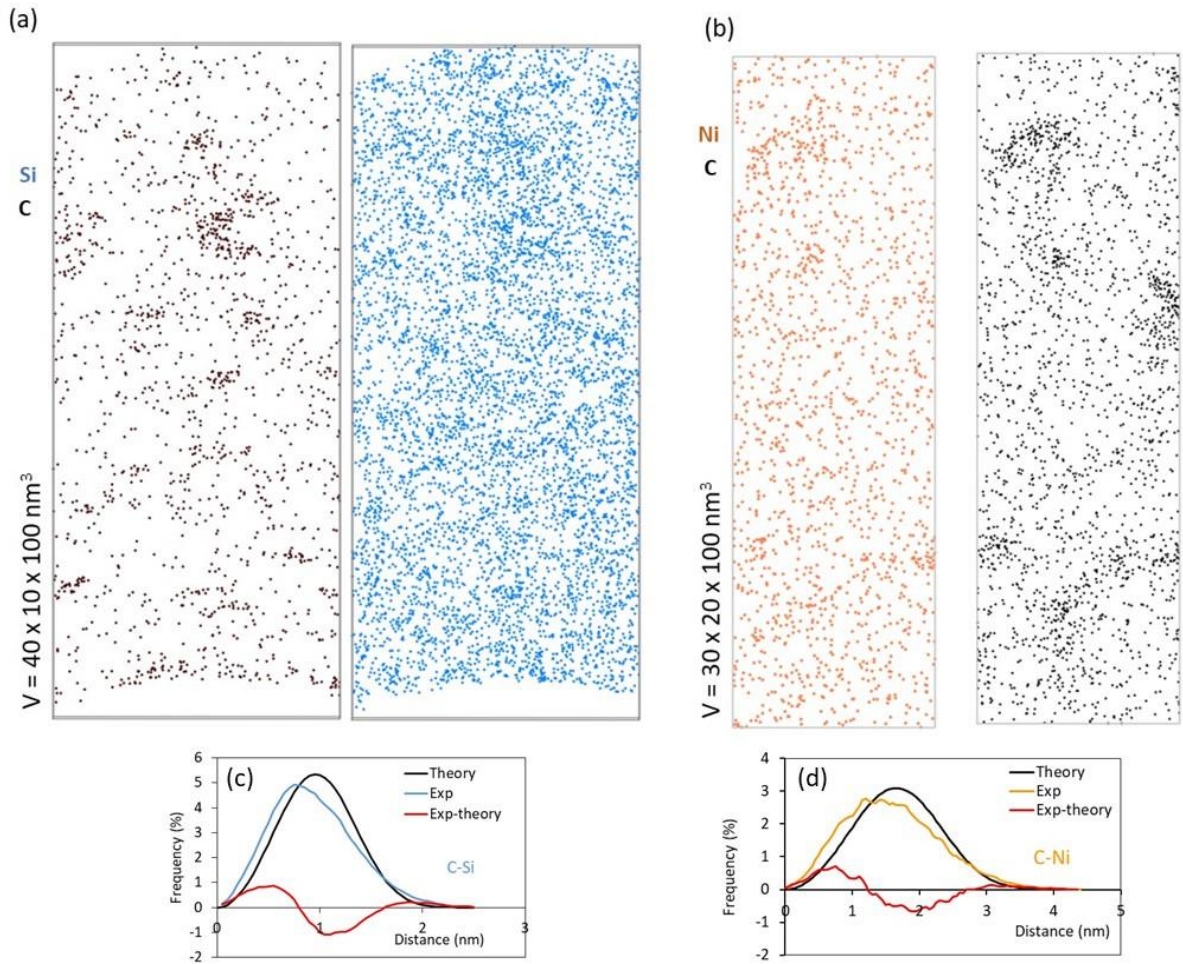

Figure 9: 3D-distributions of (a) Si and C atoms in Fe15CrSi and (b) Ni and C atoms in Fe15CrNi after irradiation at 2.5 dpa at 300 °C. 1NN and $\Delta_{1NN}$ distributions at 2.5 dpa for (c) C-Si pairs in Fe15CrSi and in (d) C-Ni pairs in Fe15CrNi.

## 4. Discussion

Clustering of Si, Ni and P has been detected in alloys containing only Ni, Si or P as well as in the alloy containing the three elements after ion irradiation. Clustering occurs at a dose as low as 0.1 dpa for all the species whatever the alloy as already reported in [25]. When the three species are present in the alloy, they form clusters altogether. Whereas this behaviour is clearly visible on 3D images of P, statistical treatment is needed to highlight the segregation of Ni and Si. $\Delta_{1NN}$ curves and composition of SRCs show that P is the specie having the strongest segregation tendency as already mentioned in [25] and predicted by modelling [14,33]. P atoms are the fastest diffusers [17,18] and present a strong positive flux coupling with vacancies and self-interstitials towards point defect sinks [14,18,33]. Because P mixed dumbbells are trapped by substitutional P atoms in the matrix [17,18], the immobilized dumbbells can act as nuclei of dislocation loops. Indeed, the experimental work of Hardouin-Duparc [35] reports the increase in number density of dislocation loops in Fe-P under irradiation. These immobile complexes become sinks that could further be enriched by Ni, Si, Cr and P [16]. This trapping and the high diffusion coefficient of P result in a high P concentration of the clusters at 0.1 dpa (Table 2). The P concentration of the SRCs decreases at 0.5 and 2.5 dpa, while the SRCs are progressively enriched by the slower diffusers, namely Ni and Si, which have also been shown to have systematic positive RIS at point defect sinks [14].

Comparison of $\Delta_{1NN}$ curves between alloys containing only one minor specie and the one obtained in the alloy containing the three of them (Figure 7) revealed that segregation level quantified by the $\Delta_{1NN}$ graphs is the same in all the alloys at 0.1 dpa as already mentioned in [25] (similar amplitudes of the peaks associated with the clusters for all the alloys). However, at 2.5 dpa, it is no more the case. At this higher dose, clustering was stronger when the three species are altogether. The trapping effect of P mixed dumbbells by P explains this observation as it favours the creation of nuclei for SRCs. However, the fact that P clustering is also stronger in Fe15CrNiSiP than in Fe15CrP shows that Ni and Si also have an influence. As a conclusion, there is a synergistic effect of the three species favouring clustering, P certainly playing a determining role.

At 2.5 dpa in Fe15CrNi and Fe15CrSi, the clusters formed by Ni and Si are enriched in C. A question arises. Would Si and Ni cluster in the absence of C? As P atoms, C atoms are known to form immobile complexes, mainly with vacancies [36,37]. Therefore, these immobile complexes become sinks that could further be enriched by Si in Fe14CrSi or by Ni in Fe14CrNi [16]. It is therefore possible that C ends up in the clusters, not because of affinity with Ni and Si, but with point defects, forming stable complexes with them, which are later enriched with Ni and Si solutes that are dragged there by single point-defects.

In Fe15CrNiSiP, both SRCs and dislocation loops enriched in Ni, Si, Cr and P were observed at 2.5 dpa. Their composition (Table 2) is very similar. This confirms that they are of similar nature. At 2.5 dpa, the number density of decorated point defect clusters is therefore equal to the sum of the number density of dislocation loops and SRCs, i.e. (4.9 ± 0.5) $10^{23}$ m$^{-3}$.

Based on these experimental results, previous work [12,25] and on the strong affinity of these species to segregate at dislocation loops and dislocation [3,7,8,14,30,35–39], the irradiation-induced nature of the SRCs as mentioned in [12,16,25] and the formation mechanism proposed in [16] are confirmed. SRCs are small decorated point defects clusters.

## 5. Conclusions

The effect of composition on SRC formation was studied based on a set of four Fe-15Cr alloys with different impurity contents (Fe15CrNiSiP, Fe15Cr-Ni, Fe15CrSi and Fe15CrP). These alloys were irradiated at 300 °C with 5 MeV Fe-ions up to 2.5 dpa.

Clustering of Ni, Si or P species has been detected from low doses (0.1 dpa) thanks to the 1NN data analysis. The clustering of P atoms is observable on the 3D atom distribution maps for all studied doses.

However, the clustering of Ni and Si species is only visually observed at high dose (2.5 dpa) but exist as shown by statistical tests.

A synergetic effect due to the presence of the three species has been evidenced, P certainly playing a determining role. The important role of P on early SRC formation is confirmed, in agreement with diffusion data [17,18] and the stronger dragging of P over the other elements as predicted by Messina et al. [14,33].

At 2.5 dpa, dislocation loops are observed and exhibit similar enrichment levels as SRCs. The clusters are first enriched by P. Ni and Si arrive later at these sinks. These results are consistent with an irradiation-induced nature of the SRCs as mentioned in [12,16,25] and the formation mechanism proposed by Castin et al. [16]

## Acknowledgements

This work has received funding from the Euratom research and training programme 2014-2018 under grant agreement No. 755039 (M4F project). This research was partly funded by the Euratom's Seventh Framework Programme FP7/2007- 2013 under grant agreement No. 604862 (MatISSE project) and contributes to the EERA (European Energy Research Alliance) Joint Programme on Nuclear Materials (JPNM). Experiments were performed on the GENESIS platform. GENESIS is supported by the Région Haute-Normandie, the Métropole Rouen Normandie, the CNRS via LABEX EMC and the French National Research Agency as a part of the program "Investissements d'avenir" with the reference ANR-11-EQPX-0020. The use of the HZDR Ion Beam Center facilities and the support by its staff is gratefully acknowledged.

## Data Availability

The raw/processed data required to reproduce these findings cannot be shared at this time as the data also forms part of an ongoing study.